\documentstyle[draft,psfig]{mn} 

\newif\ifAMStwofonts

\ifoldfss
  \ifCUPmtlplainloaded \else
    \NewTextAlphabet{textbfit} {cmbxti10} {}
    \NewTextAlphabet{textbfss} {cmssbx10} {}
    \NewMathAlphabet{mathbfit} {cmbxti10} {} 
    \NewMathAlphabet{mathbfss} {cmssbx10} {} 
  \fi
  \ifAMStwofonts
    \ifCUPmtlplainloaded \else
      \NewSymbolFont{upmath} {eurm10}
      \NewSymbolFont{AMSa} {msam10}
      \NewMathSymbol{\upi}     {0}{upmath}{19}
      \NewMathSymbol{\umu}     {0}{upmath}{16}
      \NewMathSymbol{\upartial}{0}{upmath}{40}
      \NewMathSymbol{\leqslant}{3}{AMSa}{36}
      \NewMathSymbol{\geqslant}{3}{AMSa}{3E}

    \fi
  \fi
\fi 

\ifnfssone
  \newmathalphabet{\mathit}
  \addtoversion{normal}{\mathit}{cmr}{m}{it}
  \addtoversion{bold}{\mathit}{cmr}{bx}{it}
  \newmathalphabet{\mathbfit} 
  \addtoversion{normal}{\mathbfit}{cmr}{bx}{it}
  \addtoversion{bold}{\mathbfit}{cmr}{bx}{it}
  \newmathalphabet{\mathbfss} 
  \addtoversion{normal}{\mathbfss}{cmss}{bx}{n}
  \addtoversion{bold}{\mathbfss}{cmss}{bx}{n}
  \ifAMStwofonts
    \ifCUPmtlplainloaded \else
      \UseAMStwoboldmath
      \makeatletter
      \new@mathgroup\upmath@group
      \define@mathgroup\mv@normal\upmath@group{eur}{m}{n}
      \define@mathgroup\mv@bold\upmath@group{eur}{b}{n}
      \edef\UPM{\hexnumber\upmath@group}
      \new@mathgroup\amsa@group
      \define@mathgroup\mv@normal\amsa@group{msa}{m}{n}
      \define@mathgroup\mv@bold\amsa@group{msa}{m}{n}
      \edef\AMSa{\hexnumber\amsa@group}
      \makeatother
      \mathchardef\upi="0\UPM19
      \mathchardef\umu="0\UPM16
      \mathchardef\upartial="0\UPM40
      \mathchardef\leqslant="3\AMSa36
      \mathchardef\geqslant="3\AMSa3E
    \fi
  \fi
\fi 

\ifnfsstwo
  \DeclareMathAlphabet{\mathbfit}{OT1}{cmr}{bx}{it}
  \SetMathAlphabet\mathbfit{bold}{OT1}{cmr}{bx}{it}
  \DeclareMathAlphabet{\mathbfss}{OT1}{cmss}{bx}{n}
  \SetMathAlphabet\mathbfss{bold}{OT1}{cmss}{bx}{n}
  \ifAMStwofonts
    \ifCUPmtlplainloaded \else
      \DeclareSymbolFont{UPM}{U}{eur}{m}{n}
      \SetSymbolFont{UPM}{bold}{U}{eur}{b}{n}
      \DeclareSymbolFont{AMSa}{U}{msa}{m}{n}
      \DeclareMathSymbol{\upi}{0}{UPM}{"19}
      \DeclareMathSymbol{\umu}{0}{UPM}{"16}
      \DeclareMathSymbol{\upartial}{0}{UPM}{"40}
      \DeclareMathSymbol{\leqslant}{3}{AMSa}{"36}
      \DeclareMathSymbol{\geqslant}{3}{AMSa}{"3E}
    \fi
  \fi
\fi 

\ifCUPmtlplainloaded \else
  \ifAMStwofonts \else 
    \def\upi{\pi}
    \def\umu{\mu}
    \def\upartial{\partial}
  \fi
\fi

\title[Black hole--strange star coalescence]{The swallowing of
 a quark star by a black hole}

\author[W. Klu\'{z}niak and W.~H. Lee]
{W\l odzimierz Klu\'{z}niak$^{1,2}$ and William H. Lee$^{3}$ \\
$^{1}$ Copernicus Astronomical Centre, ul. Bartycka 18, 00--716 Warszawa,
Poland \\
$^{2}$ Institute of Astronomy, Zielona G\'ora University, ul. Lubuska 2,
65--265 Zielona G\'ora, Poland\\
$^{3}$ Instituto de Astronom\'{\i}a, Universidad Nacional Aut\'{o}noma
de M\'{e}xico, Apdo. Postal 70--264, Cd. Universitaria, 04510
M\'{e}xico D.F.\\
}

\pagerange{\pageref{firstpage}--\pageref{lastpage}}
\pubyear{2002}

\begin{document}

\maketitle

\label{firstpage}


\begin{abstract}
In 3-d SPH simulations of the coalescence of a quark star with a
pseudo-Newtonian black hole all of the quark matter is quickly
accreted by the black hole. The Madsen-Caldwell-Friedman
argument against the existence of quark stars may need to be
re-examined.
\end{abstract}
   


\section{Introduction}

An astrophysical argument has been invoked against the existence of
quark stars in our Galaxy: in the coalescence of a quark star with a
comparably compact object a huge number of small fragments of quark
matter is expected to be ejected from the binary and to pollute the
galactic environment\cite{madsen88,cf91}, precluding formation of
glitching neutron stars \cite{alpar87}.  We have set out to test this
expectation against actual simulations of the coalescence process.

In a first attempt to model quark stars in smooth particle
hydrodynamics (SPH), we have already carried out strictly Newtonian
simulations of the black hole coalescence of stars modeled with an
equation of state (e.o.s.) appropriate to self-bound quark matter
\cite[henceforth paper I]{lkn01} and compared the outcome against
previously published results \cite{lk99a,lk99b,kl98,l00,l01} of
analogous simulations of the coalescence of a black hole and a
polytrope, taken to represent a neutron star.  We have found
significant differences between the two sets of simulations. Although
the star was disrupted, to a degree, and a disk of matter formed
around the black hole in each of the two cases, for the quark star
system we have found no clear evidence of mass ejection in those
Newtonian simulations, to the limit of our resolution.

Here, we report the results of 3-d hydrodynamic simulations of the
coalescence of a quark star moving in a pseudo-potential (e.g.,
Paczy\'nski and Wiita 1980) modeling salient features of general
relativistic motion around black holes.  The coalescence is over much
more quickly than in the Newtonian case. The black hole swallows the
quark star in one gulp.

\section{Quark stars} \label{quark}

If quark matter is stable at zero pressure, quark stars should exist,
and models of such stars, i.e., of a massive amount of quark fluid in
hydrostatic equilibrium, have been constructed in general relativity
\cite{itoh70,bodmer71,bc76,witten84,hzs86,afo86a}.  The properties of
rapidly rotating quark stars of all mases have also been investigated
in general relativity \cite{g99,s99,z00,gr01,a02}.

Population studies indicate that if quark stars exist at all, their
number may be comparable to the number of black holes, and a
significant number of compact binaries with a quark star member is
expected \cite{b02}.

Coalescing compact objects are prime candidates for detection with
ground based gravitational wave detectors.  Processes involving quark
stars have been invoked to explain a number of other high energy
phenomena, as well, from soft gamma repeaters to properties of radio
pulsars (e.g., Alcock, Farhi and Olinto 1986b, Horvath et al 1993,
Zhang, Xu and Qiao 2000, Usov 2001, Cheng and Dai 2002) Quark stars
are especially attractive as a theoretical candidate for a gamma ray
burst (GRB) source---models involving quark stars would neatly
sidestep the difficult problem of baryon contamination, as there are
no baryons in quark matter \cite{paczynski91}.  However, if ejection
of quark matter seeds does occur in binary coalescence of quark stars,
GRB models involving such a process \cite{h91} would be difficult to
reconcile with the presence of young neutron stars in our Galaxy
\cite{kluzniak94}.

A more complete discussion of the issues touched upon here can be
found in paper I, and in the reviews by Cheng, Dai \&
Lu~\shortcite{cdl98} and Madsen~\shortcite{madsen99}.

\section{The simulation} \label{simulation}

Quark matter can be described by a linear equation of state (e.g.,
Zdunik 2000).  In the MIT-bag model, in the limit of massless quarks,
the pressure is $P=c^2(\rho-\rho_0)/3$ \cite{fj84}, with $\rho c^2$
the energy density.  We take $2.634\times10^{14}\,{\rm g/cm^3}
<\rho_0<7.318\times10^{14}\,{\rm g/cm^3}$ in the various runs reported
here.  We numerically construct a non-rotating Newtonian star with
this e.o.s., and place it in binary orbit around a black hole, modeled
as a spherical vacuum cleaner with a pseudo-Newtonian potential
\cite{lk95,lk99b}.

The initial separation was chosen to allow a quick merger. The
spiral-in is at first caused by loss of angular momentum to
gravitational radiation.  We have found the results to be insensitive
to the initial separation and to the choice of pseudo-potential (see
below), as well as to the placement of the absorbing boundary around
the ``black hole."  The low shear viscosity of quark matter suggests
that, just as is the case for neutron stars \cite{bc92}, a quark star
in a coalescing binary will not be tidally locked. We further assume
that the star has not been born a millisecond rotator. This is why we
decided on a non-rotating quark star in the initial conditions.

We neglect the very thin crust which may be present in quark stars
\cite{afo86a}.  The star should have a sharp boundary at zero
pressure, and density $\rho_0$ (the bold contour in Fig.  1). Although
the density of self-bound quark fluid cannot drop below $\rho_0$, in
presenting the results we draw also five contours of average spatial
density of values successively lower by factors of 1.78. This allowed
us to trace the distribution of quark droplets in the Newtonian
simulation (paper I).  In the current simulation we detect no low
density regions (thin contours in Fig. 1), other than the contours
just outside the star---these are a numerical artifact reflecting the
size of the SPH kernel (close to 800 meters at the stellar boundary)
over which the spatial average is performed.

\section{Pseudo-Newtonian black hole} \label{potentials}

Paczy\'{n}ski \& Wiita~\shortcite{pw80} modeled relativistic
accretion disks using Newtonian  equations of motion in a pseudo-potential
given by
\begin{eqnarray}
\Phi_{\rm PW}(r)=-\frac{GM_{\rm BH}}{r-r_{g}},
\label{eq:PW}
\end{eqnarray}
where $r_{g}=2GM_{\rm BH}/c^{2}$.  This function does not satisfy the
Laplace equation, but is spherically symmetric, and quite useful in
reproducing such features of orbital motion around a Schwarzschild
black hole as the innermost stable orbit, and hence of modeling
relativistic accretion flows. In our simulations, we take this
potential to describe the gravity of the black hole. We place an
absorbing boundary at $r=3GM_{\rm BH}/c^{2}$, corresponding to the
radius of the photon orbit in Schwarzschild geometry.  The black disks
in Fig. 1 have this radius.  When an SPH particle touches the
boundary, it is removed from the simulation and its mass and momentum
are added to the black hole. The code conserves angular momentum.  In
other runs, not reported in Table 1, we have placed the absorbing
boundary at various other radii in the range $(2.1r_g, 4r_g)$.  There
was no change of the results.

The Paczy\'{n}ski-Wiita pseudo-potential is divergent at $r_g$, i.e.,
it is very attractive close to that radius. To test whether this
was the source of the rapid accretion of the quark star by the black hole,
we have also used another pseudo-potential devised by us to reproduce the
Schwarzschild ratio of the orbital and epicyclic frequencies: 
\begin{eqnarray}
\Phi_{\rm KL}(r)=-\frac{GM_{\rm BH}}{3r_{g}} \left(1-e^{3r_g/r}\right).
\label{eq:KL}
\end{eqnarray}
It was not.

\section{Numerical method} \label{method}

We have used smooth particle hydrodynamics for the calculations
presented here (see Monaghan 1992). The code is the same as that used
for our previous calculations of strange star--black hole coalescence
using Newtonian physics (paper I), the only difference being in the
computation of the gravitational potential produced by the black hole
and of the gravitational radiation reaction terms. For the runs
reported here, we use the pseudo-potential of eq.~\ref{eq:PW}.  Our
treatment of gravitational radiation reaction is similar to that
employed previously \cite{lk99b}, except that we now use the orbital
frequency in the Paczy\'{n}ski--Wiita potential,
$\Omega=\left(\sqrt{G(M_{\rm BH}+ M_{\rm SS})/r}\right)/(r-r_g)$,
which is larger than the Schwarzschild value.  Following Landau \&
Lifshitz~\shortcite{landau75}, we take
\begin{eqnarray} 
\frac{dE_{orb}}{dt}=-\frac{32G\mu^{2}\Omega^{6}r^{4}}{5c^{5}},
\label{eq:dedt}
\end{eqnarray}
with $\mu=M_{\rm SS}M_{\rm BH}/(M_{\rm SS}+M_{\rm BH})$ and $r$ the
binary separation.  The coalescence now proceeds much faster than in
the Newtonian runs of paper I, and the star is accreted within one
orbital period from the start of the simulation (in less than one
millisecond).

The strange star is constructed using the MIT e.o.s. for massless
quarks, as outlined in paper I.  Indeed, we use the same initial stars
as for the runs shown there. The data for our present dynamical runs
is summarized in Table~\ref{parameters}.

\begin{table*}
\caption{Basic parameters for selected runs. The table lists, for each run,
the radius and mass of the star, the initial mass ratio
$q=M_{\rm SS}/M_{\rm BH}$, the initial orbital separation, the
density of quark matter at zero pressure $\rho_{0}$, the time at which gravitational
radiation reaction was switched off, the time at which the simulation
finished, and the initial number of particles.}
 \label{parameters}
 \begin{tabular}{@{}lcccccccc}
  Run & $R_{\rm SS}$(km) & $M_{\rm SS}/M_{\odot}$ & $q$ & $r_{i}/R_{\rm SS}$ 
        & $\rho_{0}/10^{14}$g~cm$^{-3}$   
        & $t_{rad}/10^{-4}$s & $t_{f}/10^{-4}$s & $N$ \\
  A50   & 9.0 & 1.5 & 0.50 & 3.5 & 7.318 
        & 2.276 & 4.049 
        & 17,256 \\
  A30   & 9.0 & 1.5 & 0.30 & 5.0 & 7.318  
        & 3.457 & 5.311
        & 17,256 \\
  B50   & 12.0 & 2.0 & 0.50 & 3.5 & 4.116 
        & 3.038 & 5.407
        & 17,256 \\
  B30   & 12.0 & 2.0 & 0.30 & 5.0 & 4.116 
        & 4.613 & 7.088
        & 17,256 \\
  C50   & 15.0 & 2.5 & 0.50 & 3.5 & 2.634 
        & 3.800  & 6.755
        & 17,256 \\
  C30   & 15.0 & 2.5 & 0.30 & 5.0 & 2.634 
        & 5.770 & 8.863
        & 17,256 \\

 \end{tabular}

\end{table*}

\section{Results} \label{results}

In the pseudo-potentials used for the current simulation, as in
general relativity, the pull of gravity is so strong that the
centrifugal barrier has only finite height, and vanishes entirely for
angular momenta lower than that in the marginally stable orbit.  As a
result, the strange star is accreted whole by the black hole. This
happened for all the runs we performed, regardless of the initial mass
ratio, binary separation, stellar mass and radius, placement of the
black hole boundary, or the form of the pseudo-Newtonian potential
used.  No fragment of quark matter survives the encounter to form an
accretion disc around the black hole (let alone be ejected from the
system). The computed gravitational wave signal accordingly vanishes
abruptly once the star is accreted.

The results strongly suggest that in the binary coalescence of a quark
star and a black hole no matter is lost from the system at all.  No
comparable simulations have as yet been reported for two quark stars
in a binary, nor any relativistic simulation of any coalescence
process involving a quark star.  It may be assumed that the current
results are the best guide to the state of affairs: no quark nuggets
are ejected in the binary coalescence.  At present, the existence of
quark stars in our Galaxy cannot be ruled out.

\begin{figure*}
\psfig{width=\textwidth,file=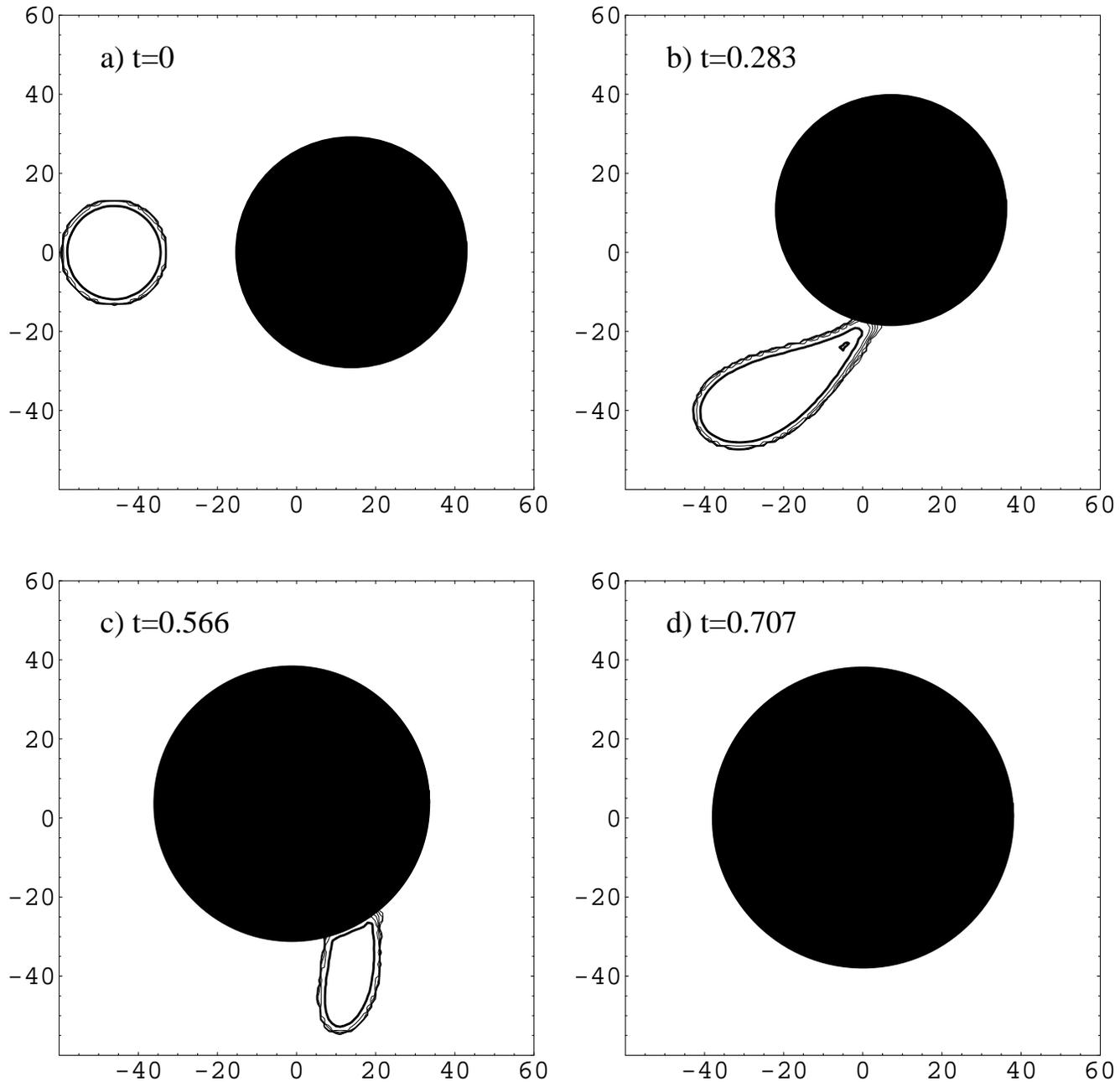,angle=0,clip=}
\caption{Density contours in the orbital plane during the dynamical
simulation of the black hole--strange star binary with initial mass
ratio $q=0.3$ and $M_{\rm SS}=2M_{\odot}$ (run B30). The orbital
rotation is counterclockwise. All contours are logarithmic and equally
spaced every 0.25 dex, with a bold contour at $\rho=\rho_{0}$. 
In fact, the presence of the thin contours is largely a numerical
artifact (Section 3). The
time for each frame is given in milliseconds, and the axes are labeled
in km.}
\label{rhoB30}
\end{figure*}

\section*{Acknowledgments}

We gratefully acknowledge the hospitality of the Institut
d'Astrophysique de Paris. WK thanks K.S. Cheng for helpful
discussions. Support for this work was provided by CONACyT (27987E)
and DGAPA--UNAM (IN-110600).

\label{lastpage}

\end{document}